\documentclass[pra,twocolumn,showpacs]{revtex4}
\usepackage{amsmath,amssymb,mathrsfs}
\usepackage{psfrag}
\usepackage{graphicx}
\usepackage{graphics}
\usepackage{epsfig}
\usepackage{bm}
\usepackage{color}
\usepackage{verbatim,color,ulem}

\newcommand{\beq}{\begin{equation}}
\newcommand{\eeq}{\end{equation}}
\newcommand{\beqa}{\begin{eqnarray}}
\newcommand{\eeqa}{\end{eqnarray}}
\newcommand{\ba}{\begin{aligned}[b]}
\newcommand{\ea}{\end{aligned}}

\newcommand{\cred}{\color{black}}
\newcommand{\cblue}{\color{black}}

\begin{document}

\title{Collisionless sound of bosonic superfluids in lower dimensions}

\author{L. Salasnich$^{1,2,3}$ {\cred and F. Sattin$^{4}$}}
\affiliation{
$^{1}$Dipartimento di Fisica e Astronomia ``Galileo Galilei'' 
and CNISM, Universit\`a di Padova, Via Marzolo 8, 35131 Padova, Italy
\\
$^{2}$Istituto Nazionale di Ottica (INO) del Consiglio 
Nazionale delle Ricerche (CNR),
via Nello Carrara 1, 50125 Sesto Fiorentino, Italy
\\
$^{3}$Istituto Nazionale di Fisica Nucleare (INFN), Sezione di Padova, 
Via Marzolo 8, 35131 Padova, Italy
\\
{\cred $^{4}$Consorzio RFX (CNR, ENEA, INFN, Universit\`a di Padova, 
Acciaierie Venete SpA), Corso Stati Uniti 4, 35127 Padova, Italy}}

\begin{abstract} 
The superfluidity of low-temperature bosons is well established 
in the collisional regime. In the collisionless regime, however, 
the presence of superfluidity is not yet fully clarified, 
in particular in lower spatial dimensions. 
Here we compare the Vlasov-Landau equation, which 
does not take into account the superfluid nature of the 
bosonic system, with the Andreev-Khalatnikov equations, 
which instead explicitly contain a superfluid velocity. 
We show that recent experimental data of the sound mode in a two-dimensional 
collisionless Bose gas of $^{87}$Rb atoms are in good agreement 
with both theories but the sound damping is better reproduced 
by the Andreev-Khalatnikov equations 
below the Berezinskii-Kosterlitz-Thouless critical temperature $T_c$ 
while above $T_c$ the {\cred Vlasov-Landau results are closer to the 
experimental ones}. For one dimensional bosonic fluids, 
where experimental data are not 
yet available, we find larger differences between the sound 
velocities predicted by the two transport theories and, also 
in this case, the existence of a superfluid velocity reduces the sound damping. 
\end{abstract}


\maketitle

\section{Introduction}

According to Landau  \cite{landau1941}, the  
liquid helium below the critical temperature is characterized 
by a superfluid component and a normal component. 
This idea was inspired by similar models {\cred used for} superconductors 
\cite{london1935} and superfluids \cite{tisza1938}. 
In the standard hydrodynamic treatment of a neutral superfluid 
\cite{ll6,khalatnikov1965,schmitt2014} the normal component is supposed 
to be in the collisional regime. 
The very special case of the collisionless superfluid Helium-4, 
where the normal component is in the collisionless regime 
was analyzed by Andreev and Khalatnikov \cite{andreev1963}. 
In the collisionless regime  \cite{khalatnikov1965,ll10} the dimensionless 
parameter $\omega \tau_c$ is such that $\omega \tau_c\gg 1$, where  $\tau_c$ 
is the collision time of quasi-particles \cite{ll10} and 
$\omega$ is the frequency 
of a generic macroscopic oscillation travelling along the fluid. 
Usually $\tau_{c}$ grows by decreasing 
the temperature $T$, and at extremely low temperature 
one expects that collisionless phenomena dominate the dynamics of 
superfluids and, more generally, the dynamics of quantum liquids. 
Indeed, the Andreev and Khalatnikov \cite{andreev1963} collisionless 
approach is in full agreement with experimental 
measurements \cite{whitney} of the sound velocity of Helium-4 
for the temperature below $0.4$ Kelvin.  
In general, depending on size and density, the system can be 
in the collisionless regime also far 
from zero temperature \cite{ll6,khalatnikov1965,schmitt2014,andreev1963,ll10}. 
Actually, natural systems as ionized plasmas 
do exist which, due to the velocity dependence of the collision 
frequency, become collisionless in the opposite regime of very 
high temperature \cite{nicholson}. 

{\cred The interest in collisionless superfluids has been renewed by a recent
experiment \cite{ville2018}, where the sound mode was measured in a
uniform quasi-2D Bose gas} made of $^{87}$Rb 
atoms. The experimental data of the speed of sound 
are in good agreement with theoretical 
results \cite{ota2018,cappellaro2018} based 
on the Vlasov-Landau equation \cite{vlasov1945,landau46} (which is  
substantially equivalent to the random-phase approximation \cite{lipparini}) 
for neutral collisionless bosons. There are, however, some 
discrepancies between the experimental data of sound damping 
and the prediction of the Vlasov-Landau equation \cite{ota2018}. 
{\cred Very recently it has been shown \cite{china} 
that the second sound of modified two-fluid hydrodynamic equations, 
which incorporate the dynamics of the quantized vortices, 
reproduce quite well the experimental sound velocity 
of Ref. \cite{ville2018}. However, in this 
dynamical Kosterlitz-Thouless theory \cite{china} 
there is a fitting parameter in the dielectric function 
which makes this theory not really predictive.} 
In Refs. \cite{ota2018,cappellaro2018} 
the superfluid nature of the system is not taken into account: 
the superfluid velocity ${\bf v}_s({\bf r},t)$ does not appear  
and the phase-space distribution $f({\bf r},{\bf p},t)$ 
of particles is used instead of the phase-space distribution 
$f_{qp}({\bf r},{\bf p},t)$ of quasi-particles. 

In this paper we investigate the collisionless sound mode 
of bosonic quantum gases both in two and one spatial dimensions. 
We compare the Vlasov-Landau equation, which 
does not take into account the superfluid nature of the 
neutral bosonic system, with the 
Andreev-Khalatnikov equations \cite{andreev1963}, 
which instead explicitly contain a superfluid velocity. 
We find that the behavior of the speed of sound 
obtained with the two approaches is similar but 
the experimental data of sound damping \cite{ville2018} 
in a 2D collisionless Bose gas are closer to the theoretical 
predictions based on Andreev-Khalatnikov equations, below 
the Berezinsky-Kosterlitz-Thouless critical temperature $T_c$ 
\cite{berezinskii1972,kosterlitz1973}. 
In 1D the superfluidity is much more elusive \cite{pitaevskii1991}, 
but it could be experimentally found at low temperature for finite-size systems 
where phase slips are inhibited \cite{super-book}. 
For the collisionless 1D Bose gas we show that 
the speed of sound predicted by the two transport theories is quite different. 
{\cred The damping rates of the sound velocities} 
are instead very close each other, but also in this 1D case the presence 
of a superfluid velocity suppresses the sound damping. 

\section{Vlasov-Landau theory of neutral collisionless bosons} 

The equilibrium distribution of a weakly-interacting gas 
of $D$-dimensional neutral bosons, each of them with mass $m$, is given by 
\beq 
f_0(p) = {1\over e^{\beta\left({p^2\over 2m} + g n_0 -\mu\right)} - 1} 
\label{eq:blv}
\eeq
where $\mu$ is the chemical potential, fixed by the condition 
$
n_0 = \int dV_{\bf p} \, f_0(p)
$
with $n_0$ the total number density at equilibrium, 
$dV_{\bf p}={d^D{\bf p}/(2\pi\hbar)^D}$ and $p=|{\bf p}|$. 
Here we assume a weakly-interacting bosonic gas 
with zero-range interaction of strength $g$. 
Notice that, because $n_0$ is constant, introducting 
the effective chemical potential ${\tilde \mu}=\mu-g n_0$, 
$f_0(p)$ can also be interpreted as the distribution of 
non-interacting bosons. 

The interaction strength $g$ appears also in the out-of-equilibrium 
mean-field external potential 
$
U_{mf}({\bf r},t) = g \int dV_{\bf p} \, f({\bf r},{\bf p},t) 
$, 
where $f({\bf r},{\bf p},t)$ is the out-of-equilibrium 
distribution function, which is driven by the following 
mean-field collisionless Vlasov-Landau equation 
\beq 
\left( {\partial \over \partial t} + {{\bf p}\over m}  
\cdot {\boldsymbol\nabla} - {\boldsymbol\nabla} U_{mf}({\bf r},t) 
\cdot {\boldsymbol\nabla}_{\bf p} \right) f({\bf r},{\bf p},t) = 0 
\label{eq:mfnlv}
\eeq
where ${\boldsymbol\nabla}=(\partial_x,\partial_y,\partial_z)$ 
and ${\boldsymbol\nabla}_{\bf p}=(\partial_{p_x},\partial_{p_y},\partial_{p_z})$. 
{\cred As previously stressed, the equibrium 
interaction term $g n_0$ is not essential 
in Eq. (\ref{eq:blv}) because it can be absorbed in the definition 
of $\mu$. Instead, the non-equilibrium interaction term $g n({\bf r},t)$ 
with $n({\bf r},t)=\int dV_{\bf p} \, f({\bf r},{\bf p},t)$ is crucial in 
the Vlasov-Landau equation (\ref{eq:mfnlv}). We observe 
that in the three-dimensional case one must 
use $2g n({\bf r},t)$ above $T_c$ 
because the exchange term in the thermal component is responsible 
for doubling the value of the density fluctuations \cite{ota2018}. 
For two-dimensional bosonic systems the absence of the factor $2$ 
is justified not only close to zero temperature but also above the 
Berezinskii-Kosterlitz-Thouless transition due to the persistence 
of a quasi-condensate regime \cite{pro2001,pro2002}.}

\subsection{Linearized Vlasov-Landau equation}

Starting from the Vlasov-Landau equation (\ref{eq:mfnlv}) and 
setting 
\beq   
f({\bf r},{\bf p},t) = f_0(p) + {\tilde f}({\bf p}) \, 
e^{i({\bf k}\cdot {\bf r}-\omega t)} 
\eeq
where {\cred $f_0(p)$ is the equilibrium distribution} 
and the plane-wave fluctuations {\cred with amplitude ${\tilde f}({\bf p})$}
are supposed to be small with respect to the equilibrium {\cred 
distribution}, we get the following linearized equation
\beq 
\left( \omega - {{\bf p}} \cdot {\bf k} \right) 
{\tilde f}({\bf p}) +  g \int dV_{{\bf p}'} {\tilde f}({\bf p}') \, 
{\bf k} \cdot {\boldsymbol\nabla}_{\bf p}f_0(p) = 0 
\label{eq:linmfblv}
\eeq
From this expression one gets an implicity formula for the 
collisionless (zero-sound) velocity $u_0=\omega/k$, namely  
\beq 
1 - g \int dV_{\bf p} { {\boldsymbol\nabla}_{\bf p}f_0(p) 
\cdot {\bf n} \over {{{\bf p}}\cdot {\bf n} - u_0}} = 0 
\label{eq:diel}
\eeq
where ${\bf n}={\bf k}/k$ with $k=|{\bf k}|$. 
Thus, linearizing Eq. (\ref{eq:mfnlv}) 
around the equilibrium configuration one obtains a plane-wave solution 
with frequency $\omega$ and wavevector ${\bf k}$ such that 
$\omega = u_0 k$, where $u_0$ is the speed of sound and $k=|{\bf k}|$. 
The determination of this complex quantity $u_0$ requires non trivial 
integrations in the complex domain of Eq. (\ref{eq:diel}) \cite{baldovin}. 
For analytical and numerical details see Appendix A. 
In general, the frequency $\omega$ and, correspondingly, 
the velocity $u_0$ are complex numbers: {\cred The real parts 
represent the actual propagation frequency/speed, 
whereas the imaginary part is the damping rate.}

\section{Andreev-Khalatnikov theory of neutral 
collisionless superfluids} 

Let us now consider a $D$-dimensional collisionless 
superfluid made of identical bosonic 
particles of mass $m$. At thermal equilibrium the system is 
characterized by the total mass density 
$
\rho_{0}=\rho_{s0}+\rho_{n0} 
\label{eq:rho}
$
where $\rho_{s0}$ is the superfluid mass density and $\rho_{n0}$ 
is the normal mass density. At fixed $\rho_0$ both $\rho_{s0}$ 
and $\rho_{n0}$ depend on the absolute temperature $T$. 
In particular, the normal mass density $\rho_{n0}$ can be 
obtained from the equilibrium distribution $f_{qp,0}(p)$ 
of quasi-particles \cite{landau1941} as 
$
\rho_{n0} = - {1\over D} 
\int dV_{\bf p} \, p^2 \, {df_0(p)\over dE} 
$ 
with $p=|{\bf p}|$ and 
\beq 
f_{qp,0}(p) = {1\over e^{\beta E[p,\rho_0]} - 1} 
\label{f0}
\eeq
where $\beta=1/(k_BT)$ with $k_B$ the Boltzmann constant 
and $E(p)$ is the spectrum 
of quasi-particles. Here we assume the Bogoliubov 
spectrum \cite{bogoliubov1947} of a weakly-interacting bosonic gas 
with zero-range interaction of strength $g$, given by 
{\cred 
\beq 
E[p,\rho_0] = \sqrt{{p^2\over 2m}\left({p^2\over 2m} + 
{2g\over m} \rho_0\right)} \;  
\label{noncicrede}
\eeq
Notice that, in the most general case, the Bogoliubov spectrum 
(\ref{noncicrede}) has a temperature dependence \cite{stoof2002}, 
which is not included in our approach.}

Within the Andreev and Khalatnikov theory 
\cite{andreev1963,khalatnikov1965,ll10}, 
the collisionless superfluid is characterized by three dynamical variables: 
the phase-space distribution of quasi-particles $f_{qp}({\bf r},{\bf p},t)$, 
the local mass density $\rho({\bf r},t)$ and the superfluid 
velocity ${\bf v}_s({\bf r},t)$. There are three coupled partial 
differential equations. One is the  collisionless Vlasov-Landau equation 
for the distribution of quasi-particles 
\beqa
\Big( {\partial \over \partial t} &+& 
{\boldsymbol\nabla}_{\bf p} \left( E[p,\rho({\bf r},t)] 
+ {\bf v}_s({\bf r},t)\cdot {\bf p} \right)
\cdot {\boldsymbol\nabla} 
- {\boldsymbol\nabla} 
\big( E[p,\rho({\bf r},t)] 
\nonumber
\\
&+& {\bf v}_s({\bf r},t)\cdot {\bf p}\big) 
\cdot {\boldsymbol\nabla}_{\bf p} \Big) f_{qp}({\bf r},{\bf p},t) = 0  
\label{vlasov}
\eeqa
where the term ${\bf v}_s({\bf r},t) \cdot {\bf p}$ in Eq. (\ref{vlasov}) 
is due to the fact that the energy of quasi-particles 
is obtained in a frame of reference at rest, in which the superfluid 
velocity is ${\bf v}_s({\bf r},t)$ \cite{ll10}. 
There is also the equation of continuity 
\beqa
{\partial\rho({\bf r},t)\over\partial t} &+& {\boldsymbol\nabla}\cdot 
\Big( \rho({\bf r},t) \, {\bf v}_s({\bf r},t) 
\nonumber
\\
&+&  \int dV_{\bf p} \, 
{\bf p} \, f_{qp}({\bf r},{\bf p},t) \Big) = 0 
\label{csf1}
\eeqa
and it is important to observe that in front of 
${\bf v}_s({\bf r},t)$ it appears $\rho({\bf r},t)$. 
Finally, there is an equation for the superfluid velocity 
${\bf v}_s({\bf r},t)$, 
which reads 
\beqa
{\partial {\bf v}_s({\bf r},t) \over\partial t} &+& {\boldsymbol \nabla} 
\Big[ {1\over 2} v_s({\bf r},t)^2 + 
{\mu_0[\rho({\bf r},t)] \over m} 
\nonumber
\\
&+& \int dV_{\bf p} \, 
{\partial E[p,\rho({\bf r},t)]\over \partial \rho} f_{qp}({\bf r},{\bf p},t) 
\Big] = {\bf 0} 
\label{csf2}
\eeqa
where 
\beq 
E[p,\rho({\bf r},t)] =  \sqrt{ {p^2\over 2m}\left( {p^2\over 2m} 
+ {2g\over m} \, \rho({\bf r},t) \right)} 
\label{super-bogoliubov}
\eeq
and $\mu_0$ the chemical potential of the system at 
zero temperature (i.e. $T=0$). 
{\cred The Landau-Vlasov equation (\ref{eq:mfnlv}) 
can be formally recovered from Eq. (\ref{vlasov}) setting  
$v_s({\bf r},t) = 0$ and expanding Eq. (\ref{super-bogoliubov}) 
for $p^2/(2m) \gg (2 g/m) \rho({\bf r},t)$. In this regime the mean-field 
force of Eq. (\ref{vlasov}) is $-\nabla E[p,\rho({\bf r},t)] 
\simeq -g \nabla n({\bf r},t)$ with $n({\bf r},t)=\rho({\bf r},t)/m$.}

\subsection{Linearized Andreev-Khalatnikov equations}

Similarly to the linearized Vlasov-Landau equation, also the linearized 
Andreev-Khalatnikov equations around the equilibrium configuration 
admit plane-wave solutions 
with frequency $\omega$ and wavevector ${\bf k}$ such that $\omega=u_0 k$ 
with $u_0$ the corresponding speed of sound. 
We linearize the Andreev-Khalatnikov equations setting 
\beqa 
f_{qp}({\bf r},{\bf p},t) &=& f_{qp,0}(p) + {\tilde f}_{qp}({\bf p}) \, 
e^{i({\bf k}\cdot {\bf r}-\omega t)}  
\\
\rho({\bf r},t) 
&=& \rho_0 + {\tilde \rho} \, e^{i({\bf k}\cdot {\bf r}-\omega t)}  
\\
{\bf v}_s({\bf r},t) &=& {\bf 0} + {\tilde {\bf v}}_s \, 
e^{i({\bf k}\cdot {\bf r}-\omega t)} 
\eeqa
where the plane-wave fluctuations are supposed to be small 
with respect to the equilibrium quantities. 
It follows that the linearized equations of motion 
are given by 
\beqa
\left( \omega - {\boldsymbol\nabla}_{\bf p} E(p) \cdot {\bf k} \right) 
{\tilde f}_{qp}({\bf p}) 
+ {\boldsymbol \nabla}_{\bf p}E(p) \cdot {\bf k} 
{df_{qp,0}(p)\over dE} 
\nonumber
\\
\left( {dE(p)\over d\rho_0} \, 
{\tilde \rho}+ {\bf p} \cdot {\tilde {\bf v}}_s \right) = 0 
\label{kha1}
\eeqa
\beq
\omega \, {\tilde \rho} - \rho_0 \, k \, {\tilde v}_s - k 
\int dV_{\bf p} \, p \, {\tilde f}_{qp}({\bf p})  = 0  
\label{kha2}
\eeq
\beqa
- \omega \, {\tilde v}_s + k \left( {1\over \rho_0} 
{dP_0\over d\rho_0} +  \int dV_{\bf p} {\tilde f}_{qp,0}({\bf p}) 
{d^2E(p)\over d\rho_0^2} \right) {\tilde\rho} 
\nonumber 
\\
+ k \int dV_{\bf p} 
{dE(p)\over d\rho_0} {\tilde f}_{qp}({\bf p}) = 0 
\label{kha3}
\eeqa
where $P_0$ is the pressure at zero temperature. 
Equations (\ref{kha2}) and (\ref{kha3}) contain respectively the terms
$\int dV_{\bf p} \, p \, {\tilde f}_{qp}$ and 
$\int dV_{\bf p} {\tilde f}_{qp,0} {d^2E(p)\over d\rho_0^2}$.
Both terms may be computed from Eq. (\ref{kha1}); thus any dependence 
from ${\tilde f}_{qp,0}$ disappears from Eqns (\ref{kha2},\ref{kha3}), 
which become a set of two linear homogeneous equations for the two 
variables ${\tilde v}_s, {\tilde \rho}$. 
The condition of vanishing determinant of the above set of linear 
equations yields the dispersion curve
\beq
\left({\cal A}- u_0 \right)^2 - ({\cal C} + c_T^2)(1 + {\cal B}) = 0 
\label{eq:determ}
\eeq
where, as before, $u_0 = \omega/k$, 
{\cblue \beq 
c_T^2 = {dP_0\over d\rho_0} + \rho_0 \int dV_{\bf p} {\tilde f}_{qp,0}({\bf p}) 
{d^2E(p)\over d\rho_0^2} 
\label{isothermal-vel}
\eeq
}
and 
\begin{eqnarray}
{\cal A} &=& \int dV_p p {\partial f_0 \over \partial p} 
{\partial E \over \partial \rho_0} { 1 \over {\partial E \over \partial p} 
- u_0} \\
{\cal B} &=& \int dV_p p^2 {\partial f_0 \over \partial p} 
{ 1 \over {\partial E \over \partial p} - u_0} \\
{\cal C} &=& \int dV_p  {\partial f_0 \over \partial p} 
\left({\partial E \over \partial \rho_0}\right)^2 
{ 1 \over {\partial E \over \partial p} - u_0} 
\label{eq:abc}
\end{eqnarray}
Analytical and numerical details on the derivation 
and solution of {\cblue Eq. (\ref{eq:determ})} are discussed in Appendix B. 

{\cred 
\section{Collisionless sound and its damping}
}

We now discuss the numerical results of the collisionless 
sound we obtain by solving the linearized Landau-Vlasov equation 
and the linearized Andreev-Khalatnikov equations. 
It is important to stress that, to investigate the 
low-temperature properties of 2D Helium 4, in Refs. 
\cite{andreev1963,khalatnikov1965,ll10} a phonon-like spectrum was used. 
Here we employ the full Bogoliubov expression.  

\begin{figure}[htbp]
\centering
\includegraphics[scale=0.7]{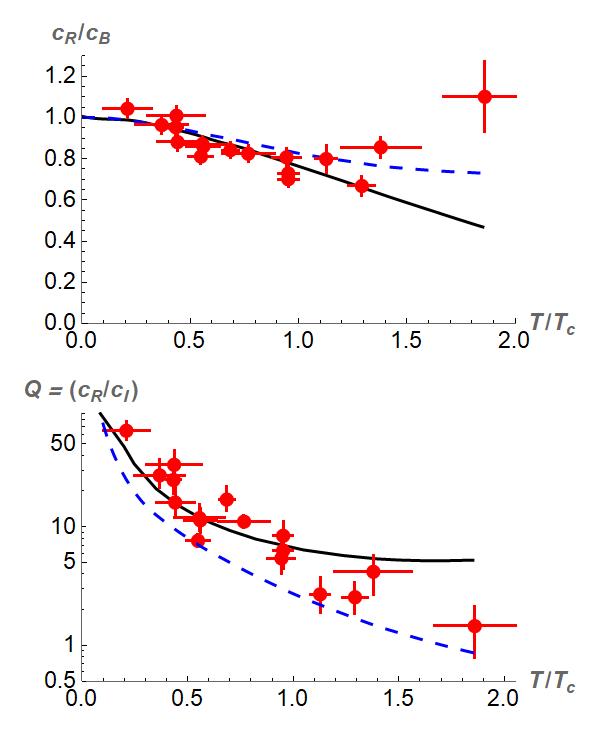}
\caption{Results from numerical solution of the dispersion equation 
$u_0 = c_R - i c_I$ vs temperature in the 2D case. 
Upper panel: the normalized speed of sound $c_R/c_B$  
as a function of the normalized temperature $T/T_c$, 
where $c_B=\sqrt{gn_0/m}$ is the Bogoliubov velocity,  
$T_c = 2\pi\hbar^2n_0/(mk_B\ln{(380\ \hbar^2/(m g))})$ is the
Berezinskii-Kosterlitz-Thouless critical temperature, and $n_0$ is the 
2D number density at equilibrium. 
Lower panel: $Q=c_R/c_I$ quality factor of the sound damping.  
To compare the two transport theories with the experiment 
of Ref. \cite{ville2018} we choose $g=0.16 \ \hbar^2/m$. 
The {\cred blue dashed} curve is the result of the Vlasov-Landau 
theory; the {\cred black solid} curve the result of the 
Andreev--Khalatnikov theory. Red dots are 
measured data of Ref. \cite{ville2018}. }
\label{fig2d}
\end{figure}

In Fig. \ref{fig2d} we report our numerical solutions  
of the speed of sound $u_0=c_R-i c_I$ in the 2D case, with $i=\sqrt{-1}$ the 
imaginary unit. {\cred Dashed} curves are obtained by using the Vlasov-Landau 
equation while {\cred solid} curves are produced 
by adopting the Andreev-Khalatnikov 
equations. In the figure there are also, as filled red circles, 
the experimental data of Ref. \cite{ville2018} obtained with a collisionless 
Bose gas of $^{87}$Rb atoms. In the figure, the quantities are 
plotted versus the scaled 
temperature $T/T_c$, with $T_c$ the Berezinskii-Kosterlitz-Thouless 
critical temperature \cite{berezinskii1972,kosterlitz1973} predicted 
at thermal equilibrium for 2D interacting superfluid bosons 
\cite{pro2001,pro2002}. The superfluid-to-normal 
Kosterlitz-Thouless phase transition occurs due to the 
unbinding of vortex-antivortex pairs, whose number strongly increases close 
to the critical temperature $T_c$. The presence of vortices with quantized 
circulation is strictly related to the existence of a superfluid 
velocity ${\bf v}_s({\bf r},t)$, which must satisfy the 
equation ${\bf v}_s({\bf r},t)=(\hbar/m){\boldsymbol\nabla}\phi({\bf r},t)$ 
with $\phi({\bf r},t)$ the angle of the phase of a complex order 
parameter \cite{stoof}. As previously stressed, the Vlasov-Landau 
equation does not include a superfluid velocity. 
Instead, the Andreev-Khalatnikov equations take into account the 
superfluid velocity but not the formation of quantized vortices 
{\cred nor the presence of a complex order parameter associated 
to the quasi-condensate \cite{super-book,pro2001,pro2002}}. 
Thus, one can expect that below $T_c$ the 2D Bose gas follows 
the Andreev-Khalatnikov while above $T_c$ the 2D bosonic system is 
better described by the Vlasov-Landau equation. 

In the upper panel of Fig. \ref{fig2d} we plot the real part of 
the scaled speed of sound $c_R/c_B$, with $c_B=\sqrt{gn_0/m}$ 
the Bogoliubov sound velocity. Remarkably, the experimental data 
(filled circles) are very well reproduced, both below and above $T_c$, 
by the Vlasov-Landau equation ({\cred dashed} curve) but also by the 
Andreev-Khalatnikov equations ({\cred solid} curve). 
At very low temperature $T$ 
the two curves of the two theories practically {\cred superimpose}. 
In the lower panel of Fig. \ref{fig2d} there is instead the 
quality factor $Q =  c_R/c_I$ of the sound damping, namely the ratio 
between the real and the imaginary part of the sound velocity $u_0=c_R-ic_I$. 
For this quality factor $Q$, the Andreev-Khalatnikov theory ({\cred 
solid} curve) is in much better agreement with 
the experimental results (filled circles) 
with respect to the Vlasov-Landau theory ({\cred dashed} curve) 
up to the critical temperature $T_c$. Above the critical temperature $T_c$ 
it seems that the quality factor $Q$ can be better reproduced by the 
Vlasov-Landau equation. {\cred Notice that in 2D the damping of the 
collisionless mode was investigated also in Ref. \cite{indy} by using 
a time-dependent Hartree-Fock-Bogoliubov approach, which practically 
gives the same results of the linearized Vlasov-Landau equation 
\cite{ota2018,lipparini}.}

\begin{figure}[htbp]
\centering
\includegraphics[scale=0.7]{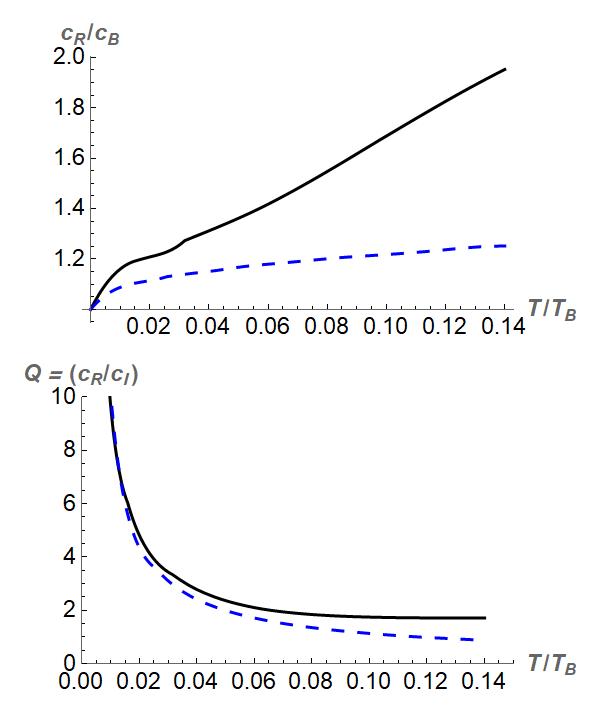}
\caption{Results from numerical solution of the dispersion 
equation $u_0 = c_R - i c_I$ {\it versus} temperature in the 1D case. 
Upper panel: the normalized speed of sound $c_R/c_B$  
as a function of the normalized temperature $T/T_B$, 
where $c_B=\sqrt{g n_0/m}$ is the Bogobliubov velocity, 
$T_B = 2 \pi n_0^2\hbar^2/(m k_B)$ is the degeneration temperature, 
and $n_0$ the 1D number density at equilibrium.  
We choose $g=0.16 \ \hbar^2n_0/m$. Lower panel: 
$Q$ quality factor of the sound 
damping: $Q = c_R/c_I$. The {\cred blue dashed} curve is the result of the 
Vlasov-Landau theory; the {\cred black solid} curve 
is obtained using the Andreev-Khalatnikov theory.}
\label{fig1d}
\end{figure}

We investigate also the 1D weakly-interacting Bose gas in 
the collisionless regime. Unfortunately there are not yet available 
experimental data in this configuration. Thus, our 1D predictions can 
be a strong benchmark for next future experiments and also 
for forthcoming theoretical investigations. 
Strictly speaking, in the thermodynamic limit and with $T>0$, for a 1D 
weakly-interacting Bose gas there is neither Bose-Einstein condensation nor 
superfluidity \cite{pitaevskii1991,super-book,stoof}. However, 
a finite 1D system of spatial size $L$ is effectively 
superfluid if $k_B T\ll E_{\phi}/ \ln{(L/\xi)}$, 
where $E_{\phi}\simeq \hbar^2n_0/(m\xi)$ is the energy needed 
to create a phase slip (topological defect, also known as black soliton) 
and $\xi=\hbar/\sqrt{2mgn_0}$ is the healing length \cite{super-book}. 

In Fig. \ref{fig1d} we show our numerical results for the complex 
speed of sound $u_0=c_R-ic_I$ of the 1D bosonic system obtained by solving 
the Vlasov-Landau equation ({\cred dashed} curves) and the Andreev-Khalatnikov 
equations ({\cred solid} curves). The quantities are plotted as a function of 
the scaled temperature $T/T_B$ where $T_B=2\pi n_0^2 \hbar^2/(m k_B)$ 
is the temperature of Bose degeneracy, where the 1D thermal 
de Broglie wavelength $\lambda_T=\hbar\sqrt{2\pi/(m k_BT)}$ becomes equal 
to the average distance $n_0^{-1}$ between bosons, with $n_0$ the 
equilibrium 1D number density. As clearly reported in the upper panel 
of Fig. \ref{fig1d}, contrary to the 2D case, in 1D the 
real part $c_R$ of the sound velocity $u_0$ increases by increasing 
the temperature $T$. However, the Andreev-Khalatnikov theory predicts 
a much larger slope. Indeed, this suggests that in 1D the determination 
of this slope can be experimentally use to the determine the superfluid 
nature of the Bose gas. {\cblue We have also found that, 
while in 2D the isothermal velocity $c_T$ of Eq. (\ref{isothermal-vel}) 
at low temperature is close to the real part of $u_0$ obtained by solving 
Eq. (\ref{eq:determ}), in the 1D system this is not the case.}

In the lower panel we plot the quality factor 
$Q=c_s/c_I$ of the sound damping: the two theoretical curves are very 
close each other. This result implies that in 1D the damping is not 
very {\cblue useful} to discriminate between the two transport theories. 

{\cred It is important to stress that marked differences 
shown in Figs. \ref{fig2d} and \ref{fig1d} are also due to the fact 
that the scaled temperature is in units of $T_c$ in 2D and 
in units of $T_B$ in 1D. 
A quite complicated analytical expression for the sound velocity $u_0$ 
of the Vlasov-Landau equation (\ref{eq:mfnlv}) can be derived 
if $f_0(p)\simeq k_BT/(p^2/2m-{\tilde\mu})$, 
i.e. under the condition $|{\tilde\mu}| \ll k_B T \ll k_B T_B$. 
In this way, in 2D one finds \cite{ota2018,cappellaro2018} that 
the real part of $u_0$ decreases by increasing in the temperature $T$, 
while in 1D we obtain the opposite, in very good agreement 
with our numerical results. As discussed in Appendix A, 
the 2D Vlasov-Landau equation can be 
reduced to an effective 1D equation but with an effective 1D Bose-Einstein 
distribution which is quite different with respect the one of the strictly  
1D case. Clearly, the behavior of $u_0$ vs $T$ crucially 
depends on the considered Bose-Einstein distribution.}

\section{Conclusions} 

We have analyzed the collisionless sound mode of a 2D weakly-interacting 
bosonic fluid, where recent experimental data are 
available \cite{ville2018}, but also the collisionless sound mode 
of the 1D bosonic fluid, where experimental data are not yet available. 
We have compared two theories: the Vlasov-Landau equation 
versus the Andreev-Khalatnikov equations. The Andreev-Khalatnikov 
equations are more sophisticated because, contrary to the 
Vlasov-Landau equation, they also take into account the presence 
of a superfluid velocity. Our 2D theoretical results, also 
confronted with the experimental data, strongly suggest that 
below the critical temperature of the superfluid-to-normal 
transition the bosonic fluid is better described by the Andreev-Khalatnikov 
theory, while above the critical temperature the Vlasov-Landau theory 
{\cred seems more} reliable. 
For the collisionless 1D Bose gas, our calculations 
show that the real part of the sound velocity grows by increasing 
the temperature and its slope determines the superfluid 
nature of the system. This prediction, as well as the reduction  
of sound damping due to the superfluid velocity, can be very useful for 
forthcoming theoretical and experimental investigations of 
collisionless superfluids. 

\section*{ACKNOWLEDGMENTS}
 
This work was partially supported by the University of Padova, BIRD project
``Superfluid properties of Fermi gases in optical potentials''. 
{\cred LS acknowledges A. Cappellaro, K. Furutani, F. Toigo, and A. Tononi 
for useful suggestions. The authors thank J. Dalibar and 
J. Beugnon for making available the experimental 
data of Ref. \cite{ville2018}.} 

\section*{Appendix A} 

In the linearized Vlasov-Landau equation (\ref{eq:diel}) 
{\cred there is the relevant quantity}  
\begin{equation}
\int d^D{\bf p} {{\nabla_p f_0}\cdot {\bf n} \over {\bf p}\cdot{\bf n} - u_0 } 
\label{eq:app1}
\end{equation}
By chosing ${\bf n}$ parallel to x-axis, this expressione simplifies to
\begin{equation}
\int d^D{\bf p} {\partial_{p_x} f_0 \over p_x - u_0 } 
\label{eq:app2}
\end{equation}
In dimension $D = 2$  it is straightforward to note that 
\begin{equation}
\int d^2{\bf p} { \partial_{p_x} f_0 \over p_x - u_0 } = \int dp_x  
\partial_{p_x}\left(\int dp_y  f_0 \right)  {1 \over p_x - u_0 }  
\label{eq:app3}
\end{equation}
Thus, both in dimension one and two, ultimately one has to deal 
with one-dimensional integrals. The integral operator comes from an 
inverse Laplace transform, hence the path of integration is defined 
in the complex $p$-plane. The recipe for choosing the right path 
was given by Landau \cite{landau46}, and may be found in several recent 
references, e.g., \cite{baldovin, nicholson}. Here we provide just 
the results. The integral (\ref{eq:app3}) writes as the sum of an 
integral along the real axis plus a contribution coming from 
poles in the complex plane:
\begin{equation}
\int_{\infty}^{+\infty} dp_x  \partial_{p_x}\left(\int dp_y  f_0 \right) 
{1 \over p_x - u_0 }  + {\cal J}
\label{eq:app4}
\end{equation}
If $Im(u_0) > 0$ then ${\cal J} = 0$. Conversely, if $Im(u_0) < 0$ we have 
\begin{equation}
{\cal J } = 2 \pi i \, \partial_{p_x} f_x(p_x = u_0) 
\label{eq:app5}
\end{equation}
with 
\begin{equation}
f_x(p_x) = \int dp_y \, p_y \ f_0(p_x,p_y) 
\end{equation}

\section*{Appendix B}

In the Andreev-Khalatnikov theory one has to deal with several 
integrals of the kind
\begin{equation}
\int dp {F(p) \over \partial_p E(p) - u_0}
\label{eq:app6}
\end{equation}
where we have dropped the $x$ lowerscript for convenience. 
$F(p)$ is one of the functions appearing in Eq. (\ref{eq:abc}). 
Since $E(p)$, as defined in (\ref{super-bogoliubov}), is 
a nonlinear function of $p$, the recipe of Eqns. 
(\ref{eq:app4},\ref{eq:app5}) needs some modifications. 
Let ${\bar p} $ be a root of the function
\begin{equation}
{\cal D}(p) = \partial_p E(p) - u_0 :  {\cal D}({\bar p}) = 0
\end{equation}
namely 
\begin{equation}
{\cal D}({\bar p}) = 0
\end{equation}
Then, we may expand ${\cal D}(p) $ around $p = {\bar p}$:
\begin{equation}
{\cal D} \simeq (p - {\bar p}) \partial_p^2 E ({\bar p})
\end{equation}
Ultimately, therefore, the integrals (\ref{eq:app6}) are evaluated as
\begin{equation}
\int dp {F(p) \over \partial_p E(p) - u_0}  = 
\int_{-\infty}^{+\infty} dp {F(p) \over \partial_p E - u_0} + {\cal J'}
\label{eq:app7}
\end{equation}   
This time we get
\begin{equation}
{\cal J'} = 2 \pi i {F({\bar p}) \over \partial_p^2 E({\bar p})}, 
\quad Im({\bar p}) < 0 
\end{equation}


\begin{thebibliography}{99}

\bibitem{landau1941} L.D. Landau, J. Phys. USSR  {\bf 5}, 71 (1941).

\bibitem{london1935} F. London and H. London, 
Proc. Royal Soc. A {\bf 149}, 866 (1935). 

\bibitem{tisza1938} L. Tisza, Nature {\bf 141}, 913 (1938).

\bibitem{ll6} L.D. Landau and E.M. Lifshitz, Fluid Mechanics, 
vol. 6 of Course of Theoretical Physics (Pergamon Press, 1987) 

\bibitem{khalatnikov1965} I.M. Khalatnikov, 
{\it An Introduction to the Theory of Superfluidity} (Pergamon Press, 1965).

\bibitem{schmitt2014} A. Schmitt, {\it Introduction to Superfluidity. 
Field-Theoretical Approach and Applications} (Springer, 2014).

\bibitem{andreev1963} A. Andreeev and I.M. Khalatnikov, 
Sov. Phys. JEPT {\bf 17}, 1384 (1963). 

\bibitem{ll10} L.D. Landau and E.M. Lifshitz, Physical Kinetics, 
vol. 10 of Course of Theoretical Physics (Butterworth-Heinemann, 1981). 

\bibitem{whitney} W.M. Whitney and C.E. Chase, Phys. Rev. Lett. 
{\bf 9}, 243 (1962). 

\bibitem{nicholson} D.R. Nicholson, {\it Introduction to plasma theory}, 
cap. 6 (Wiley, 1983). 

\bibitem{ville2018} J.L. Ville, R. Saint-Jalm, E. Le Cerf, M. Aidelsburger, 
S. Nascimbene, J. Dalibard, and J. Beugnon, 
Phys. Rev. Lett. {\bf 121}, 145301 (2018). 

\bibitem{ota2018} M. Ota, F. Larcher, F. Dalfovo, L. Pitaevskii, N.P. 
Proukakis, and S. Stringari, Phys. Rev. Lett. {\bf 121}, 145302 (2018). 

\bibitem{cappellaro2018} A. Cappellaro, F. Toigo, and L. Salasnich, 
Phys. Rev. A {\bf 98}, 043605 (2018).

\bibitem{vlasov1945} A. Vlasov, J. Phys. (Moscow) {\bf 9}, 25 (1946).

\bibitem{landau46} L.D. Landau, J. Phys. (USSR) {\bf 11}, 23 (1947).
 
\bibitem{lipparini} E. Lipparini, {\it Modern Many-Particle Physics} 
(World Scientific, 2008). 

{\cred \bibitem{china} Z. Wu, S. Zhang, and H. Zhai, 
Phys. Rev. A {\bf 102}, 043311 (2020).}

\bibitem{berezinskii1972} V. L. Berezinskii, Sov. Phys. 
JETP {\bf 34}, 610 (1972).

\bibitem{kosterlitz1973} J. M. Kosterlitz and D. J. Thouless, 
J. Phys. C: Solid State Phys. {\bf 6}, 1181 (1973).

\bibitem{pitaevskii1991} L. Pitaevskii and S. Stringari, J. Low Temp. Phys. 
{\bf 85}, 377 (1991). 

\bibitem{super-book} B. Svistunov, E. Babaev, and 
N. Prokof'ev, {\it Superfluid States of Matter} (CRC Press, Boca Raton, 2015).

\bibitem{pro2001} N. Prokof’ev, O. Ruebenacker and B. Svistunov, Phys.
Rev. Lett. {\bf 87}, 270402 (2001).

\bibitem{pro2002} N. Prokof’ev, O. Ruebenacker, 
and B. Svistunov, Phys. Rev. A {\bf 66}, 043608 (2002).

\bibitem{baldovin} F. Baldovin, A. Cappellaro, E. Orlandini, 
and L. Salasnich, J. Stat. Mech.  063303 (2016). 

\bibitem{bogoliubov1947} N. Bogoliubov, J. Phys. (USSR) {\bf 11}, 23 (1947).

{\cred 
\bibitem{stoof2002} J.O.Andersen, U. Al Khawaja, and H. T. C. Stoof, 
Phys. Rev. Lett. {\bf 88}, 070407 (2002). 
}

\bibitem{stoof} H.T.C. Stoof, K.B. Gubbels, and D.B.M. Dickerscheid,
Ultracold Quantum Fluids (Springer, 2009).

{\cred 
\bibitem{indy} M.-C. Chung and A.B. Bhattacherjee, 
New J. Phys. {\bf 11}, 123012 (2009). 
}


\end{thebibliography}
\end{document}